\documentclass[twocolumn,prl,superscriptaddress,aps,showpacs,preprintnumbers,amsmath,amssymb,a4paper]{revtex4}
\usepackage{graphicx}
\usepackage{dcolumn}
\usepackage{bm}
\usepackage[english]{babel}
\usepackage{xcolor}
\begin{document}
\preprint{APS/123-QED}
\newcommand{\hlr}[1]{{\textcolor{red}{#1}}}
\newcommand{\et}{\textit{et al.}}
\newcommand{\YCS}{YbCo$_{2}$Si$_{2}$}
\newcommand{\YRS}{YbRh$_{2}$Si$_{2}$}
\newcommand{\YCSx}{Yb(Rh$_{1-x}$Co$_{x}$)$_{2}$Si$_{2}$}
\newcommand{\TN}{$T_{N}$}
\newcommand{\TL}{$T_{L}$}
\newcommand{\YRSC}{Yb(Rh$_{0.93}$Co$_{0.07}$)$_{2}$Si$_{2}$}
\newcommand{\YRSI}{Yb(Rh$_{0.94}$Ir$_{0.06}$)$_{2}$Si$_{2}$}
\newcommand{\YRSFM}{Yb(Rh$_{0.73}$Co$_{0.27}$)$_{2}$Si$_{2}$}

\title{Doped YbRh$_2$Si$_2$: Not only Ferromagnetic Correlations but Ferromagnetic Order}

\author{S.~Lausberg}
\email{lausberg@cpfs.mpg.de}
\affiliation{Max-Planck-Institute for Chemical Physics of Solids, D-01187 Dresden, Germany}
\author{A.~Hannaske}
\affiliation{Max-Planck-Institute for Chemical Physics of Solids, D-01187 Dresden, Germany}
\author{A.~Steppke}
\affiliation{Max-Planck-Institute for Chemical Physics of Solids, D-01187 Dresden, Germany}
\author{L.~Steinke}
\affiliation{Max-Planck-Institute for Chemical Physics of Solids, D-01187 Dresden, Germany}
\author{T.~Gruner}
\affiliation{Max-Planck-Institute for Chemical Physics of Solids, D-01187 Dresden, Germany}
\author{L.~Pedrero}
\affiliation{Max-Planck-Institute for Chemical Physics of Solids, D-01187 Dresden, Germany}
\author{C.~Krellner}
\affiliation{Max-Planck-Institute for Chemical Physics of Solids, D-01187 Dresden, Germany}
\affiliation{Institute of Physics, Goethe University Frankfurt, D-60438 Frankfurt am Main, Germany}
\author{C. Klingner}
\altaffiliation[Present adress: ]{Max Planck Institute of Biochemistry, D-82152 Martinsried, Germany}
\affiliation{Max-Planck-Institute for Chemical Physics of Solids, D-01187 Dresden, Germany}
\author{M.~Brando}
\affiliation{Max-Planck-Institute for Chemical Physics of Solids, D-01187 Dresden, Germany}
\author{C.~Geibel}
\affiliation{Max-Planck-Institute for Chemical Physics of Solids, D-01187 Dresden, Germany}
\author{F.~Steglich}
\affiliation{Max-Planck-Institute for Chemical Physics of Solids, D-01187 Dresden, Germany}
\date{\today}
\begin{abstract}
YbRh$_2$Si$_2$ is a prototypical system for studying unconventional antiferromagnetic quantum criticality.
However, ferromagnetic correlations are present which can be enhanced via isoelectronic cobalt substitution for rhodium in Yb(Rh$_{1-x}$Co$_{x}$)$_{2}$Si$_{2}$.
So far, the magnetic order with increasing $x$ was believed to remain antiferromagnetic.
Here, we present the discovery of ferromagnetism for $x=0.27$ below $T_{C} = 1.30$\,K in single crystalline samples.
Unexpectedly, ordering occurs along the $c$ axis, the hard crystalline electric field direction, where the $g$ factor is an order of magnitude smaller than in the basal plane.
Although the spontaneous magnetization is only $0.1\,\mu_{B}$/Yb it corresponds to the full expected saturation moment along $c$ taking into account partial Kondo screening.
\end{abstract}
\maketitle
The unambiguous observation of antiferromagnetic (AFM) quantum critical points (QCPs) in heavy-fermion systems has led to an increasing number of theoretical and experimental works in order to understand quantum phase transitions as deeply as their classical counterparts~\cite{Stewart2001,Loehneysen2007}.
Initially, the usual model for these QCPs was the spin-density-wave scenario, but in the past ten years an alternative, the local QCP scenario, has become popular~\cite{Gegenwart2008,Si2012}.
It has been proposed to partially understand some fundamental observations, e.g. the $\omega/T$ scaling in CeCu$_{1-x}$Au$_{x}$~\cite{Schroeder2000}, but several other features remain unexplained.
Nowadays, it is clear that many of these quantum critical systems are governed by not just one single energy scale and the simple Doniach scenario with one single parameter cannot be applied~\cite{Gegenwart2007}.
Proposals concerning a global phase diagram which incorporates the effect of pressure, magnetic field and frustration on the magnetic state as well as on the Kondo effect (including Fermi surface reconstruction) have lead to a new paradigm~\cite{Senthil2004,Si2010,Coleman2010}: A system can be tuned across critical points of different nature, e.g. with itinerant or local character. In this global phase diagram the magnetic phase is considered to be antiferromagnetic.
However, real systems are more complex and often the dynamical susceptibility is governed by both finite wave vector (\textbf{q}=\textbf{Q}) and \textbf{q}=0 critical fluctuations~\cite{Gegenwart2005,Moroni-Klementowicz2009,Woelfle2011}. Theoretical calculations suggest that when a uniform (\textbf{q}=0) clean itinerant ferromagnet is tuned towards its putative ferromagnetic (FM) QCP, it becomes inherently unstable towards a phase transition of first order or modulated/textured (\textbf{q} $\neq$ 0) phases~\cite{Belitz1999,Conduit2009}. Several metals have confirmed this prediction~\cite{Suellow1999,Pfleiderer2002,Uhlarz2004,Brando2008}.

\begin{figure}[t]
\includegraphics{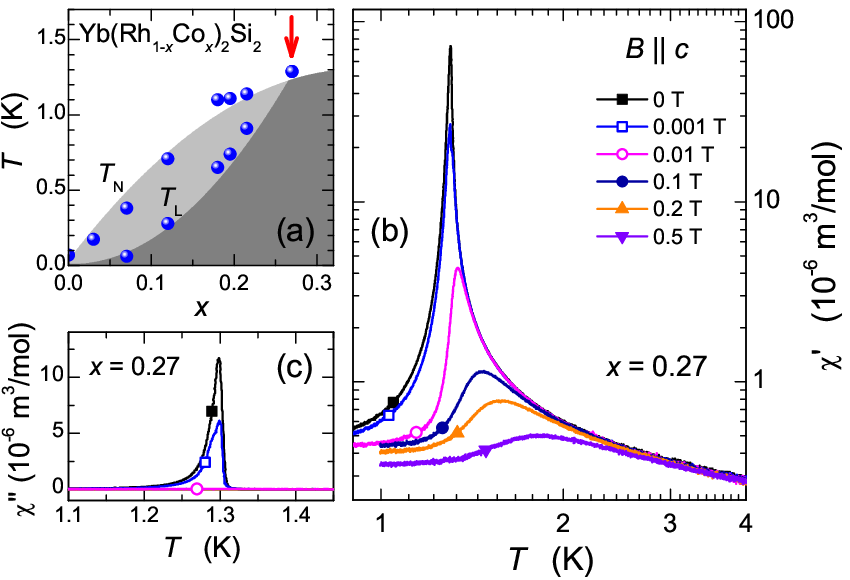}
\caption{(Color online) (a) Excerpt of the magnetic phase diagram of \YCSx\, from Ref.~\cite{Klingner2011}.
The data points correspond to the transition temperatures $T_{L}$ and $T_{N}$, respectively.
The red arrow marks the position of \YRSFM.
(b) Real part $\chi'(T)$ of the susceptibility for $x=0.27$ in different magnetic fields with $B \, \| \, c$.
At zero field a sharp peak is observed at $T_{C} = 1.30\,$K.
With increasing magnetic field the peak decreases rapidly and its position shifts towards higher temperatures.
(c) Temperature dependence of the imaginary part $\chi''(T)$.
}
\label{fig1}
\end{figure}

In this respect, the tetragonal heavy-fermion compound \YRS\, is a prototype system to study, since it displays AFM order below a very low temperature $T_{N} = 0.07$\,K~\cite{Trovarelli2000} and is very close to a peculiar AFM QCP~\cite{Custers2003}.
Its huge crystalline electric field (CEF) anisotropy is reflected by the different $g$ factor values: $g_{c} \approx 0.2$ along the $c$ axis is more than one order of magnitude smaller than $g_{a} \approx 3.6$ perpendicular to it~\cite{Gruner2012}.
Moreover, magnetic fields $B_{N, \, a} = 0.06\,$T and $B_{N, \, c} = 0.66\,$T applied perpendicular and parallel to the $c$ axis, respectively, suppress the AFM order at $T = 0$~\cite{Gegenwart2002}.
The very low $T_{N}$ and an ordered magnetic moment of $2\cdot 10^{-3}\,\mu_{B}$~\cite{Ishida2003} make it extremely difficult to determine the magnetic ordering wave vector \textbf{Q} by means of neutron scattering.
Therefore, it is helpful to increase $T_{N}$ and the magnetic moment by applying hydrostatic pressure, as this usually stabilizes magnetism in Yb systems.
This was indeed confirmed by several pressure studies~\cite{Mederle2001,Plessel2003,Tokiwa2005,Knebel2006} which in addition revealed a second phase transition at temperature $T_{L} < T_{N}$ emerging above $0.1$\,K at $0.5$\,GPa.
Comparable behavior can be obtained by chemical pressure, i.e., isoelectronic substitution of Co on the Rh site (see Fig.~\ref{fig1}a).
Good agreement of the transition temperatures was found between the pure system under hydrostatic pressure and cobalt substitution up to about $2\,$GPa, which corresponds to a Co content $x = 0.12$, ~\cite{Westerkamp2008,Friedemann2009,Klingner2011}.
At $x = 0.27$ (marked by the arrow in Fig.~\ref{fig1}a), which corresponds to a hydrostatic pressure of about $4.3\,$GPa, $T_{N}$ and $T_{L}$ appear to merge to a single magnetic transition, while in the pressure experiments by Knebel~\et\, two distinct phase transitions were observed up to $7$\,GPa~\cite{Knebel2006}.
The stabilization of the trivalent magnetic Yb state is due to a decrease of the Kondo scale under pressure.
For $x = 0.27$ a Kondo temperature $T_{K} = 7.5\,$K was deduced~\cite{Klingner2011} which is about 1/3 of that of pure \YRS.

In \YRS\, both FM and AFM fluctuations were found to coexist~\cite{Ishida2002}.
Based on field-dependent specific heat and resistivity experiments under pressure, Knebel \et\, suggested that the AFM state of \YRS\, changes into a FM one above about $5\,$GPa~\cite{Knebel2006}. Also in \YCSx\, the increase of chemical pressure increases the strength of the FM fluctuations.
This has been concluded from Curie-Weiss fits performed within a fixed temperature range above $T_{N}$, where the Weiss temperature $\Theta_{W}$ increases with increasing $x$ and even switches from negative to positive values at $x = 0.27$~\cite{Klingner2011}.
Still both phases below $T_{N}$ and $T_{L}$ were believed to be AFM since the application of a magnetic field along the magnetic easy axis ($B \, \bot \, c$) depresses the transition temperatures to zero.
This has been shown for $x=0.03$, $0.07$ and $0.12$~\cite{Westerkamp_thesis} and will be shown for $x=0.27$ ($B \, \bot \, c$) in terms of specific heat, magnetization, ac-susceptibility and electrical resistivity in a forthcoming paper~\cite{Lausberg2017}. The anisotropy in the $g$ factor becomes weaker with increasing $x$, but still persists up to \YCS.
In \YRSFM\, the value perpendicular to $c$, $g_{a} =3.2$ (at $T=5\,$K), is still about six times higher than the one parallel to $c$, $g_{c} =0.5$, while in \YCS\ it becomes $g_{a} \approx 2 \cdot g_{c}$~\cite{Gruner2012}.
Most importantly, for \YCS\ different AFM phases were found below $T_{N}$ and $T_{L}$ with the magnetic moments lying within the basal plane~\cite{Hodges1987,Klingner2011a}.

In this letter, we present the low temperature properties of \YRSFM\, by means of ac-susceptibility, specific heat and magnetization in zero field and with external magnetic fields applied mainly along the crystallographic $c$ direction. We show that the AFM ground state of \YRS\, is changed into a FM one in the Co-substituted compound with $x = 0.27$. Unexpectedly, the ordering occurs along the $c$ axis which is the CEF hard direction.

For all measurements we used the same single crystal which was grown by indium flux technique~\cite{Klingner2011}.
The low temperature measurements were performed in Oxford Instruments $^3$He-$^4$He-dilution refrigerators in a temperature range $0.02$\,K$ \leq T \leq 5$\,K and magnetic fields up to $1\,$T.
The magnetization was measured with a Faraday magnetometer~\cite{Sakakibara1994} and the ac-susceptibility with a standard susceptometer with a modulation field amplitude $B_{ac} = 15\,\mu$T. The absolute values were obtained by matching the data to measurements done in a commercial Quantum Design SQUID Vibrating Sample Magnetometer. We also used a $^{3}$He option for the SQUID (iQuantum Corporation) down to 0.5\,K. The specific heat was measured with a semi-adiabatic heat pulse technique~\cite{Wilhelm2004} and a Quantum Design Physical Properties Measurement System.

We now turn to the experimental results.
Figure \ref{fig1}b shows the real part $\chi'(T)$ of the ac-susceptibility as a function of temperature in various magnetic fields.
At zero field a sharp maximum is observed at $T_{C} = 1.30\,$K with a huge absolute value of $73 \cdot 10^{-6}\,$m$^3$/mol.
With the molar volume $V_{mol} = 46.7 \cdot 10^{-6}\,$m$^3$/mol, we obtain the dimensionless volume susceptibility $\chi_{vol} = \chi'/V_{mol} = 1.5$ which corresponds to a lower limit of the intrinsic susceptibility because of demagnetization effects due to the sample shape.
This large value of $\chi_{vol}$ points to a FM phase transition, or at least a phase transition into a magnetic structure with a FM component along the $c$ axis.
The application of an external static magnetic field $B \, \| \, c$ strongly decreases and broadens the peak.
Increasing $B$, the maximum shifts towards higher temperatures as expected for the crossover temperature of a ferromagnet.
In fig.~\ref{fig1}c the imaginary part $\chi''(T)$ of the susceptibility is depicted.
$\chi''(T)$ is large within the critical region between $1.15$ and $1.35\,$K at very low fields, presumably due to domain wall movement, characteristic of ferromagnets.
At fields above $0.001\,$T the imaginary part vanishes.

\begin{figure}[t]
\includegraphics{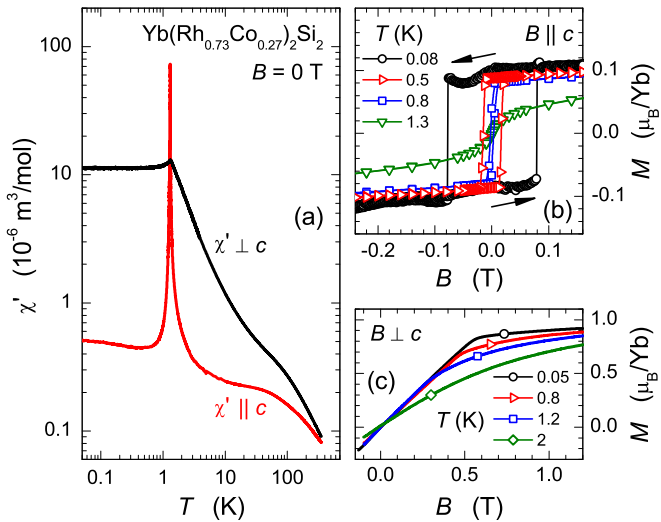}
\caption{(Color online) (a) Comparison of the susceptibility along ($\chi' \, \| \, c$) and perpendicular ($\chi' \, \bot \, c$) to the $c$ direction in zero external magnetic field.
Close to $T_{C}$, the susceptibility of the CEF hard direction $\chi' \, \| \, c$ becomes much larger than $\chi' \, \bot \, c$.
(b) Isothermal magnetization curves for different temperatures with $B \, \| \, c$.
The data were measured using different instruments.
(c) Isothermal magnetization curves for different temperatures with $B \, \bot \, c$.}
\label{fig2}
\end{figure}

Fig.~\ref{fig2}a compares the susceptibility along $c$ with its component in the basal plane $\bot \, c$.
While at temperatures above $T_{C}$ the in-plane component is much larger than $\chi' \, \| \, c$ the situation switches close to the phase transition.
The susceptibility of \YRSFM\, is reminiscent of the situation in YbNi$_4$P$_2$ which is located close to a FM QCP~\cite{Steppke2013}.
Interestingly, in both systems the magnetic hard direction, which has a tiny susceptibility at high temperatures, eventually becomes the easy direction close to $T_{C}$.
Figure \ref{fig2}b illustrates the magnetization $M(B)$ as a function of magnetic field ($B \, \| \, c$) at various temperatures.
At the lowest achieved temperature of $0.08\,$K a clear hysteresis is observed with a coercive field of about $0.08\,$T and a small spontaneous magnetic moment $0.10(2)\,\mu_{B}$/Yb.
Although on an absolute scale this spontaneous magnetization is tiny, it corresponds to $40\,$\% of the full saturation moment of the CEF ground state along $c$ as deduced from the $g$ factor determined in ESR.
Possible origins for this reduction are discussed below.
The hysteresis shrinks with increasing temperature and disappears for $T \rightarrow T_{C}$.
The $T$ dependence of the saturation magnetization roughly follows an effective $J=1/2$ Brillouin function.
Figure \ref{fig2}c shows the very different case for $B \, \bot \, c$.
Here the magnetization increases linearly with field up to a saturation value of $0.8\,\mu_{B}$/Yb at a saturation field of $B \approx 0.55\,$T.
This saturation value too corresponds to $50\,$\% of the basal plane saturation moment of the CEF ground state.
The kink in $M(B)$ related to the saturation shifts to lower fields with increasing $T$~\cite{Lausberg2017}.
Both field and $T$ dependence of the magnetization follow the expectation for a ferromagnet in a transverse field.
\begin{figure}[t]
\includegraphics{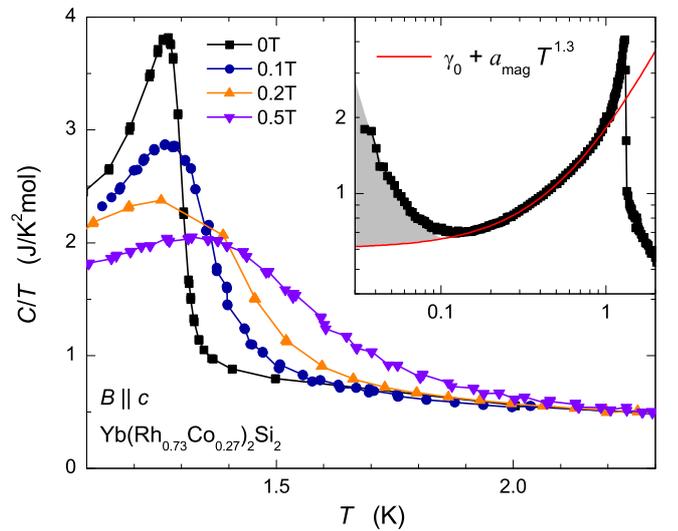}
\caption{(Color online) Specific heat $C/T$ vs $T$ of \YRSFM\, in different magnetic fields.
At $B = 0$, $C(T)/T$ features a mean-field-like anomaly at $T_{C} = 1.30\,$K.
Increasing $B$, the anomaly is shifted towards higher temperatures.
Inset: $C(T)/T$ in zero field down to $0.03\,$K.
The solid line is a fit to the magnon contribution and the shaded area depicts a nuclear contribution (see main text).
}
\label{fig3}
\end{figure}

The specific heat of \YRSFM\, is depicted in Fig.~\ref{fig3}.
At zero field, $C(T)/T$ shows a mean-field-like transition at $T_{C} = 1.30\,$K in agreement with the peak temperature found in $\chi'(T)$.
The application of a magnetic field along $c$ shifts $T_{C}$ towards higher temperatures and the maximum broadens as expected for a FM phase transition.
In the inset of Fig.~\ref{fig3} the low temperature specific heat is shown at $B = 0$.
The magnon contribution can be fitted by a power law $\gamma_0 + a_{\mathrm{mag}} \, T^{1.3}$ between $0.15$ and $1\,$K and leaves a Sommerfeld coefficient $\gamma_{0}=0.61$\,J/K$^{2}$mol~\cite{Comment03}.
Below $0.15\,$K, $C(T)/T$ strongly increases with decreasing temperature because of a contribution $C_{\mathrm{n}}(T)/T \propto T^{-3}$ (shaded area)~\cite{Steppke2010}, which is mainly caused by nuclear quadrupole and dipole contributions of the isotopes $^{171}$Yb and $^{173}$Yb.
The dipole contribution of both isotopes should vanish in zero magnetic field.
In the present case, however, it does not because the ordered $4f$ local moments induce a splitting of the Yb nuclear dipoles.
This nuclear dipole contribution can be used to get an estimate of the ordered moment.
With the hyperfine coupling constant for Co~\cite{Comment01} and Yb~\cite{Plessel2003}, we can calculate the size of the nuclear dipole moment~\cite{Steppke2010}.
Here the contribution due to the electric nuclear quadrupole moment is estimated from a linear interpolation between the \YCS\,\cite{Hodges1987} and \YRS\,\cite{Knebel2006} M\"ossbauer measurements.
The resulting ordered $4f$ magnetic moment $m$ extracted from $C_{\mathrm{n}}(T)/T$ is $0.19(6)\,\mu_{B}$/Yb, for $m$ parallel to the $c$ axis and $0.23(6)\,\mu_{B}$/Yb for $m$ perpendicular to $c$.
The total ordered moment obtained from the nuclear dipole contribution is slightly larger than the observed spontaneous ferromagnetic moment and cannot rule out a canted AFM state.
This would imply from the measured ferromagnetic $c$ component and the estimate from the nuclear specific heat a staggered in-plane moment of about $0.18(8)\,\mu_{B}$/Yb.
However, for a canted AFM state one would expect a decrease of the ordering temperature for field applied along the $c$ direction, in contrast to our experimental observation.
Furthermore, in a standard canted AFM system application of a field along the weak FM component leads to a further increase of the magnetization because of further canting in contrast to the saturation observed here.
More importantly, as discussed below, the observed spontaneous magnetization along the $c$ direction corresponds to the full unscreened part of the CEF ground state $c$ moment, while an in-plane staggered moment of $0.18\,\mu_{B}$/Yb would correspond to less than $25\,$\% of the unscreened part of the CEF ground state in-plane moment.
Therefore this would correspond to a full ferromagnetic order of the $c$ component, but only a very partial AFM ordering of the in-plane component.
Thus even for a possible canted state the ordering is essentially ferromagnetic along the $c$ direction.

\begin{figure}[t]
\includegraphics{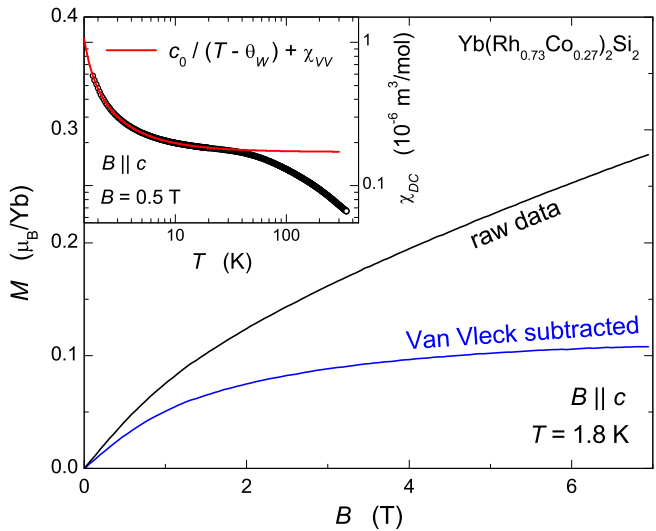}
\caption{(Color online) Magnetization as a function of magnetic field $B \, \| \, c$ at $T = 1.8\,$K. The Van Vleck contribution
$\chi_{VV} = 0.172 \cdot 10^{-6}\,$m$^3$/mol was subtracted from the raw data, leaving a saturation magnetization
at 7\,T of about $0.11\,\mu_{B}$/Yb. Inset: data of $\chi_{DC} = M/B$ between $2$ and $300\,$K.
Below $20\,$K a Curie-Weiss fit (solid line) was carried out as described in the main text.
}
\label{fig4}
\end{figure}

There is another route to obtain more information about the size of the ordered moment along $c$.
In Fig.~\ref{fig4} the magnetization $M$ is plotted as a function of $B$ above $T_{C}$ at $T = 1.8\,$K (upper curve).
Up to a field of $7\,$T the curve does not saturate which is due to a considerable CEF-induced Van Vleck contribution.
To determine this contribution we measured $\chi_{DC} = \mu_0\,M/B$ as a function of $T$ in an external magnetic field of $0.5\,$T.
The data are shown in the inset of Fig.~\ref{fig4}.
The broad shoulder above $20\,$K corresponds to the first excited doublet of the $J = 7/2$ multiplet of the Yb$^{3+}$ atom split by the crystalline electric field.
We have fitted the region below $20\,$K with the function $\chi(T) = c_0/(T - \Theta_{W}) + \chi_{VV}$ where $c_0$ is the Curie's constant, $\Theta_{W} = 1.24\,$K is the Weiss temperature and $\chi_{VV} = 0.172 \cdot 10^{-6}\,$m$^3$/mol is the constant Van Vleck term.
In the main panel of Fig.~\ref{fig4} we have subtracted the Van Vleck contribution $\chi_{VV}$ from $M(B)$ and obtained the lower curve which seems to saturate above $7\,$T at about $0.11\,\mu_{B}$/Yb. 
This value confirms that the observed spontaneous magnetization corresponds to the maximum available unscreened part of the local moment along the $c$ direction.
However this value is about a factor of $2$ smaller than the value $0.25(5)\,\mu_{B}$/Yb for the saturation moment along $c$ of the CEF ground state as deduced from the $g$ factor observed in ESR.
As pointed out above, the observed saturation moment in the basal plane is also a factor of $2$ smaller than the expected CEF ground state moment.
An obvious origin for this reduction is a partial screening due to the Kondo effect.
This is confirmed by an analysis of the entropy connected with the magnetic ordering.
It can be easily obtained by integrating the measured $C(T)/T$ after subtraction of the nuclear part and $\gamma_0$.
We get a value of about $0.21\,$R\,$\ln2$, which is a factor of at least four below the value expected for ordering in a 3D localized effective $S = 1/2$ system, in agreement with the ratio $T_{K}/T_{C}$.
From this ratio one would expect a stronger screening of the moment than just by a factor of $2$.

To conclude, we have shown that \YRSFM\, is a ferromagnet with a Curie temperature $T_{C} = 1.30$\,K and a remanent magnetization along the crystallographic $c$ direction. This is evidenced by the huge absolute values of the ac-susceptibility, the hysteresis in the magnetization isotherms at $T < T_{C}$, and by the evolution of the transition in magnetic field.
The low temperature magnetization hysteresis provides a spontaneous magnetic moment of $0.10(2)\,\mu_{B}$/Yb pointing along the $c$ direction in the FM phase.
The analysis of the low $T$ magnetization indicates that the saturation magnetization along both directions is reduced by a factor of about two compared to the values of the CEF ground state, very likely due to partial Kondo screening.
This implies that the observed spontaneous magnetization corresponds to the full expected saturation moment along the $c$ direction, i.e., full FM order of the $c$ component.
In contrast, analysis of the nuclear specific heat shows that a staggered moment in the basal plane can amount to at most $25\,$\% of the expected in-plane saturation moment.
Thus a very large part of the in-plane moment is definitively not ordered.
Therefore it is appropriate to consider \YRSFM\, as a dominantly ferromagnetic system.
Our results indicate that upon starting from \YCS\, and substituting Rh for Co, the direction of the ordered moment switches from in-plane to out-of-plane, although the ratio between out-of-plane to in-plane CEF ground state moment decreases by a factor of $3$.

Our discovery of ferromagnetism in \YRSFM\, opens the question whether the low-lying phase below $T_{L}$ which is found for small Co concentrations in \YCSx\, or in pure \YRS\, (under pressure) is ferromagnetic, too.
This might imply a field-induced FM QCP at $B_{L}<B_{N}$.
Hence, our study urges the reinvestigation of the magnetic order in \YCSx\, with $x < 0.27$ and in pure \YRS\, under small pressures with field along the $c$ direction.
The strong competition between the AFM and FM correlations is a key to understand the peculiar behavior of \YRS\, as well as the spin-liquid state in \YRSI\,~\cite{Friedemann2009}.

We are indebted to O. Stockert and S. Wirth for useful discussions as well as to C. Klausnitzer for experimental support. Part of the work was supported by the DFG Research Unit 960 Quantum Phase Transitions.


\begin{thebibliography}{10}
\bibitem{Comment03}
This power law cannot be used to discriminate between FM and AFM magnons. The
  expected anisotropy gap shall significantly influence the magnetic
  excitations in this $T$ range.
\end{thebibliography}
\end{document}